\documentclass[aps,twocolumn,showpacs]{revtex4}
\usepackage{graphicx}
\include{amssym}

\begin{document}
.
\vspace{0.4cm}

\title{Light Quark Masses in Multi-Quark Interactions} 

\author{A. A. Osipov\footnote{Email address: osipov@nu.jinr.ru}, 
        B. Hiller\footnote{Email address: brigitte@teor.fis.uc.pt}
    and A. H. Blin\footnote{Email address: alex@teor.fis.uc.pt}}
\affiliation{Centro de F\'{\i}sica Computacional, Departamento de
         F\'{\i}sica da Universidade de Coimbra, 3004-516 Coimbra, 
         Portugal}

\begin{abstract}
We suggest and discuss in detail a multi-quark three flavor Lagrangian of the 
Nambu -- Jona-Lasinio type, which includes a set of effective interactions 
proportional to the current quark masses. It is shown that within the dynamical 
chiral symmetry breaking regime, the masses of the pseudo Goldstone bosons and 
their chiral partners, members of the low lying scalar nonet, are in perfect 
agreement with current phenomenological expectations. The role of the new 
interactions is analyzed. 
\end{abstract}

\maketitle








\section{Introduction}

A long history of applying the Nambu -- Jona-Lasinio (NJL) model in hadron 
physics shows the importance of the concept of effective multi-quark 
interactions for modelling QCD at low energies. Originally formulated only in 
terms of four-fermion chiral-symmetric couplings \cite{Nambu:1961,Vaks}, the 
model has been extended to the realistic three flavor and color case with 
$U(1)_A$ breaking six-quark 't Hooft interactions \cite{Hooft:1976,Hooft:1978,Bernard:1988,Bernard:1988a,Reinhardt:1988,Weise:1990,Vogl:1990,Weise:1991,Takizawa:1990,Klevansky:1992,Hatsuda:1994,Bernard:1993,Dmitrasinovic:1990,Naito:2003} 
and a set of eight-quark interactions \cite{Osipov:2005b}. The last ones 
complete the number of vertices which are important in four dimensions for 
dynamical chiral symmetry breaking \cite{Andrianov:1993a,Andrianov:1993b}. 

The explicit breaking of chiral symmetry in the model is described by the quark
mass term of the QCD Lagrangian, e.g. \cite{Ebert:1986,Bijnens:1993}. As a 
result, deviations from the exact symmetry predictions are expressed by 
functions of the light quark masses. The current quark mass dependence is of 
importance for several reasons, in particular for the phenomenological 
description of meson spectra and meson-meson interactions, and for the critical
point search in hot and dense hadronic matter. In the latter case it has a 
strong impact on the phase diagram. The mass effects may lead to a different 
phase structure. For instance, the large mass difference between $s$ and $u 
(d)$ quarks may disfavor the formation of the color-flavor-locked phase at 
intermediate density, and the conjecture regarding the two critical points 
structure finally may not be true \cite{Kunihiro:2010}. 

The explicit chiral symmetry breaking (ChSB) by the standard mass term of the 
free Lagrangian is only a part of the more complicated picture arising in 
effective models beyond leading order \cite{Gasser:1982}. Chiral perturbation 
theory \cite{Weinberg:1979,Pagels:1975,Gasser:1984,Gasser:1985} gives a 
well-known example of a self consistent accounting of the mass terms, order by
order, in an expansion in the masses themselves. In fact, NJL-type models 
should not be an exception from this rule. If one considers multi-quark 
effective vertices, to the extent that 't Hooft and eight-quark terms are 
included in the Lagrangian, certain mass dependent multi-quark interactions 
must be also taken into account. It is the purpose of this paper to study such 
higher order terms in the quark mass expansion. In particular, we show the 
ability of the model with new quark-mass-dependent interactions to describe the
spectrum of the pseudo Goldstone bosons, including the fine tuning of the 
$\eta\!-\!\eta'$ splitting, and the spectrum of the light scalar mesons: 
$\sigma (600)$, $\kappa (850)$, $f_0(980)$, and $a_0(980)$. 

There are several motivations for this work. In the first place, the quark 
masses are the only parameters of the QCD Lagrangian which are responsible for 
an explicit ChSB, and it is important for the effective theory to trace this 
dependence in full detail. In this paper it will be argued that it is from the 
point of view of the $1/N_c$ expansion that the new quark mass dependent 
interactions must be included in the NJL-type Lagrangian already when the 
$U(1)_A$ breaking 't Hooft determinantal interaction is considered. This point 
is somehow completely ignored in the current literature. 

A second reason is that nowadays it is getting clear that the eight-quark 
interactions, which are almost inessential for the mesonic spectra in the 
vacuum, can be important for the quark matter in a strong magnetic background 
\cite{Hiller:2007,Gatto:2010,Gatto:2011,Frasca:2011,Gatto:2012}. We will show 
that there is a set of the effective quark-mass-dependent interactions which 
are of importance here and have not been considered yet. 

A further motivation comes from the hadronic matter studies in a hot and dense 
environment. It is known that lattice QCD at finite density suffers from 
the numerical sign problem. Thus, the phase diagram is notoriously difficult 
to compute ``ab initio'', except for an extremely high density regime where 
perturbative QCD methods are applicable. In such circumstances effective models
designed to shed light on the phase structure of QCD are valuable, especially 
if such models are known to be successful in the description of the hadronic 
matter at zero temperature and density. Reasonable modifications of the NJL 
model are of special interest in this context and our work aims at future 
applications in that area. 

\section{Effective multi-quark interactions}

The chiral quark Lagrangian has predictive power for the energy range which is 
of order $\Lambda\simeq 4\pi f_\pi\sim 1$\ GeV \cite{Georgi:1984}. $\Lambda$ 
characterizes the spontaneous chiral symmetry breaking scale. Consequently, the
effective multi-quark interactions, responsible for this dynamical effect, are
suppressed by $\Lambda$, which provides a natural expansion parameter in the 
chiral effective Lagrangian. The scale above which these interactions disappear
and QCD becomes perturbative enters the NJL model as an ultraviolet cut-off for 
the quark loops. Thus, to build the NJL type Lagrangian we have only three 
elements: the quark fields $q$, the scale $\Lambda$, and the external sources 
$\chi$, which generate explicit symmetry breaking effects -- resulting in mass 
terms and mass-dependent interactions. 

The color quark fields possess definite transformation properties with respect 
to the chiral flavor $U(3)_L\times U(3)_R$ global symmetry of the QCD 
Lagrangian with three massless quarks (in the large $N_c$ limit). It is 
convenient to introduce the $U(3)$ Lie-algebra valued field $\Sigma =(s_a-ip_a)
\frac{1}{2}\lambda_a$, where $s_a=\bar q\lambda_aq$, $p_a=\bar q\lambda_ai
\gamma_5q$, and $a=0,1,\ldots ,8$, $\lambda_0=\sqrt{2/3}\times 1$, $\lambda_a$ 
being the standard $SU(3)$ Gell-Mann matrices for $1\leq a \leq 8$. Under 
chiral transformations: $q'=V_Rq_R+V_Lq_L$, where $q_R=P_R q, q_L=P_Lq$, and 
$P_{R,L}=\frac{1}{2}(1\pm\gamma_5)$. Hence, $\Sigma'=V_R\Sigma V_L^\dagger$, and 
$\Sigma^{\dagger'}=V_L\Sigma^\dagger V_R^\dagger$. The transformation property of 
the source is supposed to be $\chi' =V_R\chi V_L^\dagger$.  
 
Any term of the effective multi-quark Lagrangian without derivatives can be 
written as a certain combination of fields which is invariant under chiral 
$SU(3)_R\times SU(3)_L$ transformations and conserves $C, P$ and $T$ discrete
symmetries. These terms have the general form
\begin{equation}
\label{genL}
   L_i\sim \frac{\bar g_i}{\Lambda^\gamma}\chi^\alpha\Sigma^\beta,
\end{equation} 
where $\bar g_i$ are dimensionless coupling constants (starting from eq. 
(\ref{h}) the dimensional couplings $g_i=\bar g_i/\Lambda^{\gamma}$ will be also
considered). Using dimensional arguments we find $\alpha+3\beta -\gamma =4$,
with integer values for $\alpha, \beta$ and $\gamma$. 

We obtain a second restriction by considering only the vertices which make 
essential contributions to the gap equations in the regime of dynamical chiral 
symmetry breaking, i.e. we collect only the terms whose contributions to the 
effective potential survive at $\Lambda\to\infty$. We get this information by 
contracting quark lines in $L_i$, finding that this term contributes to the 
power counting of $\Lambda$ in the effective potential as $\sim 
\Lambda^{2\beta-\gamma}$, i.e. we obtain that $2\beta-\gamma \geq 0$ (we used 
the fact that in four dimensions each quark loop contributes as $\Lambda^2$). 

Combining both restrictions we come to the conclusion that only vertices with
\begin{equation}
\label{ineq}
   \alpha +\beta \leq 4
\end{equation}
must be taken into account in the approximation considered. On the basis of 
this inequality one can conclude that (i) there are only four classes of 
vertices which contribute at $\alpha=0$; those are four, six and eight-quark 
interactions, corresponding to $\beta=2,3$ and $4$ respectively; the $\beta=1$ 
class is forbidden by chiral symmetry requirements; (ii) there are only six 
classes of vertices depending on external sources $\chi$, they are: $\alpha 
=1, \beta =1,2,3$; $\alpha =2, \beta =1,2$; and $\alpha =3, \beta =1$.

Let us consider now the structure of multi-quark vertices in detail. The 
Lagrangian corresponding to the case (i) is well known
\begin{eqnarray}
\label{L-int}
   L_{int}&=&\frac{\bar G}{\Lambda^2}\mbox{tr}\left(\Sigma^\dagger\Sigma\right)
   +\frac{\bar\kappa}{\Lambda^5}\left(\det\Sigma+\det\Sigma^\dagger\right) 
   \nonumber \\
   &+&\frac{\bar g_1}{\Lambda^8}\left(\mbox{tr}\,\Sigma^\dagger\Sigma\right)^2
   +\frac{\bar g_2}{\Lambda^8}\mbox{tr}
   \left(\Sigma^\dagger\Sigma\Sigma^\dagger\Sigma\right).
\end{eqnarray}  
It contains four dimensionful couplings $G, \kappa, g_1, g_2$.

The second group (ii) contains eleven terms
\begin{equation}
   L_\chi =\sum_{i=0}^{10}L_i,
\end{equation}
where
\begin{eqnarray}
\label{L-chi-1}
   L_0&=&-\mbox{tr}\left(\Sigma^\dagger\chi +\chi^\dagger\Sigma\right)
   \nonumber \\
   L_1&=&-\frac{\bar\kappa_1}{\Lambda}e_{ijk}e_{mnl}
   \Sigma_{im}\chi_{jn}\chi_{kl}+h.c.
   \nonumber \\
   L_2&=&\frac{\bar\kappa_2}{\Lambda^3}e_{ijk}e_{mnl}
   \chi_{im}\Sigma_{jn}\Sigma_{kl}+h.c.
   \nonumber \\
   L_3&=&\frac{\bar g_3}{\Lambda^6}\mbox{tr}
   \left(\Sigma^\dagger\Sigma\Sigma^\dagger\chi\right)+h.c.
   \nonumber \\
   L_4&=&\frac{\bar g_4}{\Lambda^6}\mbox{tr}\left(\Sigma^\dagger\Sigma\right)
   \mbox{tr}\left(\Sigma^\dagger\chi\right)+h.c.
   \nonumber \\
   L_5&=&\frac{\bar g_5}{\Lambda^4}\mbox{tr}\left(\Sigma^\dagger\chi
   \Sigma^\dagger\chi\right)+h.c.
   \nonumber \\
   L_6&=&\frac{\bar g_6}{\Lambda^4}\mbox{tr}\left(\Sigma\Sigma^\dagger\chi
   \chi^\dagger +\Sigma^\dagger\Sigma\chi^\dagger\chi\right)
   \nonumber \\
   L_7&=&\frac{\bar g_7}{\Lambda^4}\left(\mbox{tr}\Sigma^\dagger\chi 
   + h.c.\right)^2
   \nonumber \\ 
   L_8&=&\frac{\bar g_8}{\Lambda^4}\left(\mbox{tr}\Sigma^\dagger\chi 
   - h.c.\right)^2
   \nonumber \\ 
   L_9&=&-\frac{\bar g_9}{\Lambda^2}\mbox{tr}\left(\Sigma^\dagger\chi
   \chi^\dagger\chi\right)+h.c.
   \nonumber \\
   L_{10}&=&-\frac{\bar g_{10}}{\Lambda^2}\mbox{tr}\left(\chi^\dagger\chi\right)
   \mbox{tr}\left(\chi^\dagger\Sigma\right)+h.c.
\end{eqnarray}
Each term in the Lagrangian $L_6$ is hermitian by itself, but because of the 
parity symmetry of strong interactions, which transforms one of them into the 
other, they have a common coupling $\bar g_6$. 

Some useful insight into the Lagrangian above can be obtained by considering it
from the point of view of the $1/N_c$ expansion. Indeed, the number of color 
components of the quark field $q^i$ is $N_c$, hence summing over color indices 
in $\Sigma$ gives a factor of $N_c$, i.e. one counts $\Sigma\sim N_c$. 

The cut-off $\Lambda$ that gives the right dimensionality to the multi-quark 
vertices scales as $\Lambda\sim N_c^0=1$. On the other hand, since the leading 
quark contribution to the vacuum energy is known to be of order $N_c$, the 
first term in (\ref{L-int}) is estimated as $N_c$, and we conclude that $G\sim 
1/N_c$. 

Furthermore, the $U(1)_A$ anomaly contribution (the second term in 
(\ref{L-int})) is suppressed by one power of $1/N_c$, it yields $\kappa\sim 
1/N_c^3$. 

The last two terms in (\ref{L-int}) have the same $N_c$ counting as the 't 
Hooft term. They are of order $1$. Indeed, Zweig's rule violating effects
are always of order $1/N_c$ with respect to the leading order contribution 
$\sim N_c$. This reasoning helps us to find $g_1\sim 1/N_c^4$. The term with 
$g_2\sim 1/N_c^4$ is also $1/N_c$ suppressed. It represents the next to the 
leading order contribution with one internal quark loop in $N_c$ counting. 
Such vertex contains the admixture of the four-quark component $\bar qq\bar qq$
to the leading quark-antiquark structure at $N_c\to\infty$.  

Next, all terms in eq. (\ref{L-chi-1}), except $L_0$, are of order 1. The 
argument is just the same as before: this part of the Lagrangian is obtained 
by succesive insertions of the $\chi$-field ($\chi$ counts as $\chi\sim 1$) in 
place of $\Sigma$ fields in the already known $1/N_c$ suppressed vertices. It 
means that $\kappa_1, g_9, g_{10}\sim 1/N_c$, $\kappa_2, g_5, g_6, g_7, 
g_8\sim 1/N_c^2$, and $g_3, g_4\sim 1/N_c^3$.     

There are two important conclusions here. The first is that at leading order 
in $1/N_c$ only two terms contribute: the first term of eq. (\ref{L-int}), and 
the first term of eq. (\ref{L-chi-1}). This corresponds exactly to the standard
NJL model picture, where mesons are pure $\bar qq$ states. At the next to 
leading order we have thirteen terms additionally. They trace the Zweig's rule 
violating effects $(\kappa, \kappa_1, \kappa_2, g_1, g_4, g_7, g_8, g_{10})$, and
an admixture of the four-quark component to the $\bar qq$ one ($g_2, g_3, g_5, 
g_6$, $g_9$). Only the phenomenology of the last three terms 
from eq. (\ref{L-int}) has been studied until now. We must still understand 
the role of the other ten terms to be consistent with the generic $1/N_c$ 
expansion of QCD. 
 
The second conclusion is that the $N_c$ counting justifies the classification 
of the vertices made above on the basis of the inequality (\ref{ineq}). This is
seen as follows: the equivalent inequality $\lceil (\alpha +\beta )/2\rceil
\leq 2$ is obtained by restricting the multi-quark Lagrangian to terms that do
not vanish at $N_c\to\infty$ (it follows from (\ref{genL}) that $\beta -\lceil
\gamma /2\rceil\geq 0$ by noting that $\bar g_i\sim 1/N_c^{\lceil\gamma /2\rceil}$,
where $\lceil\gamma /2\rceil$ is the nearest integer greater than or equal to 
$\gamma /2$). 

The total Lagrangian is the sum    
\begin{equation}
\label{LQ}
   L=\bar qi\gamma^\mu\partial_\mu q+L_{int}+L_\chi. 
\end{equation}
In this $SU(3)_L\times SU(3)_R$ symmetric chiral Lagrangian we neglect terms 
with derivatives in the multi-quark interactions, as usually assumed in the
NJL model. We follow this approximation, because the specific questions for
which these terms might be important, e.g. the radial meson excitations, or 
the existence of some inhomogeneous phases, characterized by a spatially 
varying order parameter, are not the goal of this work.

Finally, having all the building blocks conform with the symmetry content of 
the model, one is now free to choose the external source $\chi$. Putting $\chi 
={\cal M}/2$, where 
$$
   {\cal M}=\mbox{diag}(\mu_u, \mu_d, \mu_s),
$$ 
we obtain a consistent set of explicitly breaking chiral symmetry terms. This 
leads to the following mass dependent part of the NJL Lagrangian    
\begin{equation}
   L_\chi \to L_\mu= -\bar qmq+\sum_{i=2}^8L_i'
\end{equation}
where the current quark mass matrix $m$ is equal to
\begin{eqnarray}
\label{cqmm}
   m&=&{\cal M}+\frac{\bar\kappa_1}{\Lambda}\left(\det {\cal M}
   \right){\cal M}^{-1}+\frac{\bar g_9}{4\Lambda^2}{\cal M}^3 
   \nonumber \\
   &+&\frac{\bar g_{10}}{4\Lambda^2}\left(\mbox{tr}{\cal M}^2\right){\cal M},
\end{eqnarray}
and
\begin{equation}
\label{L-chi-2}
   \begin{array}{lcr}
   L_2'=\frac{\bar\kappa_2}{2\Lambda^3}e_{ijk}e_{mnl}
   {\cal M}_{im}\Sigma_{jn}\Sigma_{kl}+h.c.
   \\ \\
   L_3'=\frac{\bar g_3}{2\Lambda^6}\mbox{tr}
   \left(\Sigma^\dagger\Sigma\Sigma^\dagger{\cal M}\right)+h.c.
   \\ \\
   L_4'=\frac{\bar g_4}{2\Lambda^6}\mbox{tr}\left(\Sigma^\dagger\Sigma\right)
   \mbox{tr}\left(\Sigma^\dagger{\cal M}\right)+h.c.
   \\ \\
   L_5'=\frac{\bar g_5}{4\Lambda^4}\mbox{tr}\left(\Sigma^\dagger{\cal M}
   \Sigma^\dagger{\cal M}\right)+h.c.
   \\ \\
   L_6'=\frac{\bar g_6}{4\Lambda^4}\mbox{tr}\left[{\cal M}^2
   \left(\Sigma\Sigma^\dagger +\Sigma^\dagger\Sigma\right)\right]
   \\ \\
   L_7'=\frac{\bar g_7}{4\Lambda^4}\left(\mbox{tr}\Sigma^\dagger{\cal M} 
   + h.c.\right)^2
   \\ \\ 
   L_8'=\frac{\bar g_8}{4\Lambda^4}\left(\mbox{tr}\Sigma^\dagger{\cal M} 
   - h.c.\right)^2
   \end{array}
\end{equation}

Let us note that there is a definite freedom in the definition of the external 
source $\chi$. In fact, the sources 
\begin{eqnarray}
   \chi^{(c_i)}&=&\chi +\frac{c_1}{\Lambda}
   \left(\det\chi^\dagger\right)\chi\left(\chi^\dagger\chi\right)^{-1}+
   \frac{c_2}{\Lambda^2}\chi\chi^\dagger\chi  
   \nonumber \\
   &+&\frac{c_3}{\Lambda^2}\mbox{tr}\left(\chi^\dagger\chi\right)\chi
\end{eqnarray} 
with three independent constants $c_i$ have the same symmetry transformation
property as $\chi$. Therefore, we could have used $\chi^{(c_i)}$ everywhere 
that we used $\chi$. As a result, we would come to the same Lagrangian with 
the following redefinitions of couplings 
\begin{eqnarray}
\label{rt}
   &&\bar\kappa_1\to\bar\kappa_1'=\bar\kappa_1+\frac{c_1}{2}, \quad 
   \bar g_5\to\bar g_5'=\bar g_5-\bar\kappa_2c_1,
   \nonumber \\
   &&\bar g_7\to\bar g_7'=\bar g_7+\frac{\bar\kappa_2}{2}c_1,  \quad   
   \bar g_8\to\bar g_8'=\bar g_8 +\frac{\bar\kappa_2}{2}c_1, \quad
   \nonumber \\
   &&\bar g_9\to\bar g_9'=\bar g_9 +c_2 -2\bar\kappa_1 c_1, \quad 
   \nonumber \\
   &&\bar g_{10}\to\bar g_{10}'=\bar g_{10}+c_3+2\bar\kappa_1 c_1.    
\end{eqnarray}
Since $c_i$ are arbitrary parameters, this corresponds to a continuous family 
of equivalent Lagrangians. This family reflects the known Kaplan -- Manohar 
ambiguity \cite{Manohar:1986,Leutwyler:1990,Donoghue:1992,Leutwyler:1996} in 
the definition of the quark mass, and means that several different parameter 
sets (\ref{rt}) may be used to represent the data. In particular, without loss 
of generality we can use the repara\-metrization freedom to obtain the set 
with $\bar\kappa_1' =\bar g_9' =\bar g_{10}'=0$. 

The effective multi-quark Lagrangian can be written now as  
\begin{equation}
\label{qlagr}
   L=\bar q(i\gamma^\mu\partial_\mu -m)q+L_{int}+\sum_{i=2}^8L_i'. 
\end{equation} 
It contains eighteen parameters: the scale $\Lambda$, three parameters which
are responsible for explicit chiral symmetry breaking $\mu_u,\mu_d,\mu_s$, 
and fourteen interaction couplings $\bar G, \bar\kappa, \bar\kappa_1, \bar
\kappa_2$, $\bar g_1,\ldots,\bar g_{10}$. Three of them, $\bar\kappa_1, 
\bar g_9, \bar g_{10}$, contribute to the current quark masses $m$. Seven more 
describe the strength of multi-quark interactions with explicit symmetry 
breaking effects. These vertices contain new details of the quark dynamics 
which have not been studied yet in any NJL-type models.   

\section{From quarks to mesons: stationary phase calculations}

The model can be solved by path integral bosonization of this quark Lagrangian.
Indeed, following \cite{Reinhardt:1988} we may equivalently introduce auxiliary
fields $s_a=\bar q\lambda_aq,\, p_a=\bar qi\gamma_5\lambda_aq$, and physical 
scalar and pseudoscalar fields $\sigma =\sigma_a\lambda_a,\,\phi = \phi_a
\lambda_a$. In these variables the Lagrangian is a bilinear form in quark 
fields (once the replacement has been done the quarks can be integrated out 
giving us the kinetic terms for the physical fields $\phi$ and $\sigma$)
\begin{eqnarray}
\label{L}
   L&\!=\!&\bar q\left[i\gamma^\mu\partial_\mu -\left(\sigma + i\gamma_5\phi
   \right)\right]q + L_{aux}, \nonumber \\
   L_{aux}&\!=\!&s_a\sigma_a + p_a\phi_a - s_am_a +L_{int}(s,p) \nonumber \\
   &\!+\!&\sum_{i=2}^8L_i'(s,p,\mu).
\end{eqnarray}   
It is clear, that after the elimination of the fields $\sigma,\,\phi$ by means 
of their classical equations of motion, one can rewrite this Lagrangian in its 
original form (\ref{qlagr}). On the other hand, written in terms of auxiliary 
bosonic variables, the Lagrangian becomes
\begin{eqnarray}
   L_{int}(s,p)&\!=\!&L_{4q}+L_{6q}+L_{8q}^{(1)}+L_{8q}^{(2)},
   \nonumber \\ 
   L_{4q}(s,p)&\!=\!&\frac{\bar G}{2\Lambda^2}\left(s_a^2+p_a^2\right),
   \nonumber \\
   L_{6q}(s,p)&\!=\!&\frac{\bar\kappa}{4\Lambda^5}A_{abc}s_a(s_bs_c-3p_bp_c),
   \\
   L_{8q}^{(1)}(s,p)&\!=\!&\frac{\bar g_1}{4\Lambda^8}\left(s_a^2+p_a^2\right)^2,
   \nonumber \\ 
   L_{8q}^{(2)}(s,p)&\!=\!&\frac{\bar g_2}{8\Lambda^8}\left[d_{abe}d_{cde}
   \left(s_as_b+p_ap_b\right)\left(s_cs_d+p_cp_d\right)\right.
   \nonumber \\ 
   &\!+\!&\left.4f_{abe}f_{cde}s_as_cp_bp_d\right], \nonumber
\end{eqnarray}  
and the quark mass dependent part is as follows
\begin{eqnarray}
\label{L-chi-3}
   L_2'&\!=\!&\frac{3\bar\kappa_2}{2\Lambda^3}A_{abc}\mu_a\left(s_bs_c-p_bp_c
   \right), \nonumber \\
   L_3'&\!=\!&\frac{\bar g_3}{4\Lambda^6}\mu_a
   \left[d_{abe}d_{cde}s_b\left(s_cs_d+p_cp_d\right)-2f_{abe}f_{cde}p_bp_cs_d
   \right],
   \nonumber \\
   L_4'&\!=\!&\frac{\bar g_4}{2\Lambda^6}\mu_bs_b\left(s_a^2+p_a^2\right),
   \nonumber \\
   L_5'&\!=\!&\frac{\bar g_5}{4\Lambda^4}\mu_b\mu_d\left(d_{abe}d_{cde}-
   f_{abe}f_{cde}\right)\left(s_as_c-p_ap_c\right),
   \nonumber \\
   L_6'&\!=\!&\frac{\bar g_6}{4\Lambda^4}\mu_a\mu_bd_{abe}d_{cde}
   \left(s_cs_d+p_cp_d\right),
   \nonumber \\
   L_7'&\!=\!&\frac{\bar g_7}{\Lambda^4}\left(\mu_as_a\right)^2, 
   \nonumber \\ 
   L_8'&\!=\!&-\frac{\bar g_8}{\Lambda^4}\left(\mu_ap_a\right)^2,
\end{eqnarray}
where 
\begin{equation}
   A_{abc}=\frac{1}{3!}e_{ijk}e_{mnl}(\lambda_a)_{im}(\lambda_b)_{jn}
   (\lambda_c)_{kl},
\end{equation} 
and the $U(3)$ antisymmetric $f_{abc}$ and symmetric $d_{abc}$ constants are 
standard. 

Our final goal is to clarify the role of the mass-dependent terms described by 
the Lagrangian densites of eq. (\ref{L-chi-3}). We can gain some understanding 
of this by considering the low-energy meson dynamics which follows from our 
Lagrangian. For that we must exclude quark degrees of freedom in (\ref{L}), 
e.g., by integrating them out from the corresponding generating functional. 
The standard Gaussian path integral leads us to the fermion determinant, which 
we expand by using a heat-kernel technique 
\cite{Osipov:2006a,Osipov:2001,Osipov:2001a,Osipov:2001b}. The remaining part 
of the Lagrangian, $L_{aux}$, depends on auxiliary fields which do not have 
kinetic terms. The equations of motion of such a static system are the 
extremum conditions 
\begin{equation}
\label{sp}
   \frac{\partial L}{\partial s_a}=0, \quad \frac{\partial L}{\partial p_a}=0, 
\end{equation}  
which must be fulfilled in the neighbourhood of the uniform vacuum state of the 
theory. To take this into account one should shift the scalar field $\sigma\to
\sigma +M$. The new $\sigma$-field has a vanishing vacuum expectation value 
$\langle\sigma\rangle =0$, describing small amplitude fluctuations about the 
vacuum, with $M$ being the mass of constituent quarks. We seek solutions of eq. 
(\ref{sp}) in the form:
\begin{eqnarray}
\label{st}
   s_a^{st}&=&h_a+ h_{ab}^{(1)}\sigma_b + h_{abc}^{(1)}\sigma_b\sigma_c
   +h_{abc}^{(2)}\phi_b\phi_c + \ldots
   \nonumber \\
    p_a^{st}&=&h_{ab}^{(2)}\phi_b + h_{abc}^{(3)}\phi_b\sigma_c +\ldots
\end{eqnarray}
Eqs. (\ref{sp}) determine all coefficients of this expansion giving rise to a 
system of cubic equations to obtain $h_a$, and the full set of recurrence 
relations to find higher order coefficients in (\ref{st}). We can gain some
insight into the physical meaning of these parameters if we calculate the
Lagrangian density $L_{aux}$ on the stationary trajectory. In fact, using the
recurrence relations, we are led to the result
\begin{eqnarray}
\label{lam}
   L_{aux}\!\!\!\!\!\!
   &&=h_a\sigma_a+\frac{1}{2}\,h_{ab}^{(1)}\sigma_a\sigma_b  
      +\frac{1}{2}\,h_{ab}^{(2)}\phi_a\phi_b  \\
   &&+\,\frac{1}{3}\,\sigma_a\left[h^{(1)}_{abc}\sigma_b\sigma_c
   +\left(h^{(2)}_{abc}+h^{(3)}_{bca}\right)\phi_b\phi_c\right]
   + \ldots \nonumber
\end{eqnarray} 
From this one can see that $h_a$ define the quark condensates, $h_{ab}^{(1)}$,
$h_{ab}^{(2)}$ contribute to the masses of scalar and pseudoscalar states, and
higher order $h$'s are the couplings that measure the strength of the 
meson-meson interactions.

We proceed now to explain the details of determining $h$. In the 
following only the first coefficients $h_a$, $h_{ab}^{(1)}$, and $h_{ab}^{(2)}$ 
will be of interest to us. In particular, eq. (\ref{sp}) states that $h_a=0$, 
if $a\neq 0,3,8$, while $h_\alpha$ ($\alpha =0,3,8$), after the convenient 
redefinition to the flavor indices $i=u,d,s$ 
\begin{equation} 
   h_\alpha =e_{\alpha i} h_i, \quad
   e_{\alpha i}=\frac{1}{2\sqrt 3}\left(
          \begin{array}{ccc}
          \sqrt 2&\sqrt 2&\sqrt 2 \\
          \sqrt 3&-\sqrt 3& 0 \\
          1&1&-2 
          \end{array} \right),
\end{equation}
satisfy the following system of cubic equations
\begin{eqnarray}
\label{h}
   &&\Delta_i + \frac{\kappa}{4}t_{ijk}h_jh_k +\frac{h_i}{2}\left(
     2G+g_1h^2 + g_4\mu h\right) + \frac{g_2}{2}h_i^3 \nonumber \\
   &&+\frac{\mu_i}{4}\left[3g_3h_i^2 +g_4h^2 +2(g_5+g_6)\mu_ih_i 
     +4g_7 \mu h\right] \nonumber \\
   &&+\kappa_2t_{ijk}\mu_jh_k=0.
\end{eqnarray}
Here $\Delta_i=M_i-m_i$; $t_{ijk}$ is a totally symmetric quantity, whose 
nonzero components are $t_{uds}=1$; there is no summation over the open index 
$i$ but we sum over the dummy indices, e.g. $h^2=h_u^2+h_d^2+h_s^2, 
\mu h=\mu_uh_u+\mu_dh_d+\mu_sh_s$. 

In particular, eq. (\ref{cqmm}) reads in this basis
\begin{equation}
\label{cqm-2}
   m_i=\mu_i\left(1+\frac{g_9}{4}\mu_i^2+\frac{g_{10}}{4}\mu^2\right)
   +\frac{\kappa_1}{2}t_{ijk}\mu_j\mu_k.
\end{equation}
For the set $g_9=g_{10}=\kappa_1=0$ the current quark mass $m_i$ coincides 
precisely  with the explicit symmetry breaking parameter $\mu_i$. 

Note that the factor multiplying $h_i$ in the third term of eq. (\ref{h}) is 
the same for each flavor. This quantity also appears in all meson mass 
expressions, and there is no further dependence on the couplings $G, g_1, g_4$
involved for meson states with $a=1,\ldots,7$. Thus there is a freedom of 
choice which allows to vary these couplings, condensates and quark masses 
$\mu_i$, without altering this part of the meson mass spectrum. 

It is now straightforward to obtain the inverse matrices to $h_{ab}^{(1)}$ and 
$h_{ab}^{(2)}$, namely
\begin{eqnarray}
\label{h1}
   &&-2\left(h_{ab}^{(1)}\right)^{-1}= \left(2G+g_1h^2+g_4\mu h\right)\delta_{ab}
   +4g_1h_ah_b \nonumber \\
   &&+3A_{abc}\left(\kappa h_c+2\kappa_2\mu_c\right) 
   +g_2h_{r}h_{c}\left(d_{abe}d_{cre}+2d_{ace}d_{bre}\right) 
   \nonumber \\
   &&+g_3\mu_{r}h_{c}\left(d_{abe}d_{cre}+d_{ace}d_{bre}+d_{are}d_{bce}
   \right) \nonumber \\
   &&+2g_4\left(\mu_ah_b+\mu_bh_a\right)
   +g_5\mu_r\mu_c\left(d_{are}d_{bce}-f_{are}f_{bce}\right)
   \nonumber \\
   &&+g_6\mu_r\mu_cd_{abe}d_{cre}+4g_7\mu_a\mu_b.
\end{eqnarray}
\begin{eqnarray}
\label{h2}
   &&-2\left(h_{ab}^{(2)}\right)^{-1}= \left(2G+g_1h^2+g_4\mu h\right)\delta_{ab}
   \nonumber \\
   &&-3A_{abc}\left(\kappa h_c+2\kappa_2\mu_c\right)
   +g_2h_{r}h_{c}\left(d_{abe}d_{cre}+2f_{are}f_{bce}\right) 
   \nonumber \\
   &&+g_3\mu_{r}h_{c}\left(d_{abe}d_{cre}+f_{are}f_{bce}+f_{ace}f_{bre}
   \right) \nonumber \\
   &&-g_5\mu_r\mu_c\left(d_{are}d_{bce}-f_{are}f_{bce}\right)
   \nonumber \\
   &&+g_6\mu_r\mu_cd_{abe}d_{cre}-4g_8\mu_a\mu_b.
\end{eqnarray}
These coefficients are totally defined in terms of $h_a$ and the parameters of 
the model. 

\section{From quarks to mesons: heat kernel calculations}

We now turn our attention to the total Lagrangian of the bosonized theory. To
write down this Lagrangian we should add the terms coming from integrating out 
the quark degrees of freedom in (\ref{L}) to our result (\ref{lam}). 
Fortunately, the result is known. One can find all necessary details of such 
calculations for instance in \cite{Osipov:2006a}, where we used the modified 
heat kernel technique \cite{Osipov:2001,Osipov:2001a,Osipov:2001b} developed 
for the case of explicit chiral symmetry breaking. Here we quote the main 
outcome. The $\sigma$ tadpole term must be excluded from the total Lagrangian. 
This gives us a system of gap equations. 
\begin{equation}
\label{gap}
   h_i+\displaystyle\frac{N_c}{6\pi^2} M_i 
   \left[3I_0-\left(3M_i^2-M^2\right) I_1 \right]=0. 
\end{equation}
Here $N_c=3$ is the number of colors, and $M^2=M_u^2+M_d^2+M_s^2$. The factors 
$I_i$ $(i=0,1,\ldots )$ are the arithmetic average values $I_i=\frac{1}{3}
[J_i(M_u^2)+J_i(M_d^2)+J_i(M_s^2)]$, constructed from the one-quark-loop 
integrals 
\begin{equation}
\label{ji}
   J_i(m^2)=\int\limits_0^\infty\frac{{\rm d}t}{t^{2-i}}\rho 
   (t\Lambda^2) e^{-t m^2},
\end{equation}
with the Pauli-Villars regularization kernel \cite{Osipov:2004a,Osipov:2004b}
\begin{equation}
   \rho (t\Lambda^2)=1-(1+t\Lambda^2)\exp (-t \Lambda^2).
\end{equation}
In the following we need only to know two of them
\begin{equation}
\label{j0}
   J_0(m^2)=\Lambda^2- m^2\ln\left(1+\frac{\Lambda^2}{m^2}\right),
\end{equation}
and
\begin{equation}
\label{j1}
      J_1(m^2)=\ln\left(1+\frac{\Lambda^2}{m^2}\right)  
      -\frac{\Lambda^2}{\Lambda^2+m^2}\ .
\end{equation}

From now on we will consider the case with an exact $SU(2)$ isospin symmetry, 
i.e. $\mu_u=\mu_d=\hat\mu\neq \mu_s$, and $M_u=M_d=\hat M\neq M_s$. We also 
restrict ourselves to small perturbations, so we retain terms in the bosonized
Lagrangian which are quadratic in the perturbations $\phi$ and $\sigma$. To 
this order we obtain
\begin{eqnarray}
\label{mass}
     L&\!=\!&\frac{N_cI_1}{16\pi^2}\,\mbox{tr}\left[
     (\partial_\mu \sigma )^2+(\partial_\mu \phi )^2\right] 
     +\frac{N_cI_0}{4\pi^2}\left(\sigma_a^2+\phi_a^2\right) \nonumber \\
     &\!-\!&\frac{N_cI_1}{12\pi^2}
     \left\{
     \Delta_{ns}\left[2\sqrt{2}(3\sigma_0\sigma_8+\phi_0\phi_8)
     -\phi_8^2+\phi_i^2\right]
     \right.\nonumber \\
     &\!+\!& 2(2\hat M^2+M_s^2)\sigma_0^2+(\hat M^2+5M_s^2)
     \sigma_8^2 \nonumber \\
     &\!+\!&(7\hat M^2-M_s^2)\sigma_i^2+(\hat M+M_s)(\hat M+2M_s)
     \sigma_f^2 \nonumber \\
     &\!+\!&\left.(M_s-\hat M)(2M_s-\hat M)\phi_f^2\right\} 
     \nonumber \\ 
     &\!+\!&\frac{1}{2}\,h_{ab}^{(1)}\sigma_a\sigma_b 
     +\frac{1}{2}\,h_{ab}^{(2)}\phi_a\phi_b +\ldots ,
\end{eqnarray}
where $\Delta_{ns}=\hat M^2-M_s^2$, $\phi^2_i=\sum_{i=1}^3\phi_i^2$,
$\phi^2_f=\sum_{f=4}^7\phi_f^2$. The kinetic term requires a redefinition
of meson fields, 
\begin{equation}
   \sigma_a =g\sigma_a^R, \quad \phi_a =g\phi_a^R, \quad 
   g^2=\frac{4\pi^2}{N_cI_1},
\end{equation}
to obtain the standard factor $1/4$. The Lagrangian (\ref{mass}) in the chiral
limit, $m=0$, leads to the conserved vector, ${\cal V}_\mu^a$, and 
axial-vector, ${\cal A}_\mu^a$, currents. The matrix elements of axial-vector 
currents 
\begin{equation}
      \langle 0|{\cal A}_\mu^a (0)|\phi^b_R(p)\rangle = ip_\mu f^{ab}
\end{equation}
define the weak and electromagnetic decay constants of physical pseudoscalar 
states (see details in \cite{Osipov:2006a}). In fact, we obtain that all new 
information about the mass-dependent interactions is explicitly absorbed in 
the last two terms of the Lagrangian, where the matrices $h_{ab}^{(1,2)}$ are 
block diagonal and mix only in the $(0,8)$ sector, see eqs. (\ref{h1}) and 
(\ref{h2}). There is also an implicit dependence through the gap and 
stationary phase equations.    

\section{Fixing parameters}

Now let us fix the values of the various quantities introduced. After choosing 
the set $\kappa_1=g_9=g_{10}=0$ we still have to fix fourteen parameters: 
$\Lambda, \hat m, m_s, G, \kappa, \kappa_2$ and $g_1,\ldots,g_8$. Note that 
there are two intrinsic restrictions of the model, namely, the stationary phase
(\ref{h}) and the gap (\ref{gap}) equations, which must be solved 
selfconsistently. This is how the explicit symmetry breaking is intertwined with
the dynamical symmetry breaking and vise versa.  
We use (\ref{gap}) to determine $\hat h, h_s$ through 
$\Lambda, M_s$ and $\hat M$. The ratio $M_s/\hat M$ is related to the ratio of 
the weak decay constants of the pion, $f_\pi=92$\ MeV, and the kaon, 
$f_K=113$\ MeV. Here we obtain 
\begin{equation}
   \frac{M_s}{\hat M}=2\frac{f_K}{f_\pi}-1=1.46. 
\end{equation} 
Furthermore, the two eqs. (\ref{h}) can be used to find the values of 
$\Lambda$ and $\hat M$ if the parameters $\hat m$, $m_s$, $G$, $\kappa$, 
$\kappa_2$, $g_1, \ldots, g_7$ are known. Thus, together with $g_8$ we have at 
this stage thirteen couplings to be fixed. Let us consider the current quark 
masses $\hat m$ and $m_s$ to be an input. Their values are known, from various 
analyses of the chiral treatment of the light pseudoscalars, to be around 
$\hat m=4$\ MeV and $m_s=100$\ MeV \cite{PDG}. Then the remaining eleven 
couplings can be found by comparing with empirical data. One should stress the 
possibility (which did not exist before the inclusion of mass-dependent 
interactions) to fit the low lying pseudoscalar spectrum, $m_\pi=138$\ MeV, 
$m_K=494$\ MeV, $m_\eta=547$\ MeV, $m_{\eta'}=958$\ MeV, the weak pion and kaon 
decay constants, $f_\pi=92$\ MeV, $f_K=113$\ MeV, and the singlet-octet mixing 
angle $\theta_p=-15^\circ$ to perfect accuracy. One can deduce that the 
couplings $\kappa_2$ and $g_8$ are essential to improve the description in the 
pseudoscalar sector; in particular, $g_8$ is responsible for fine tuning the 
$\eta\!-\!\eta'$ mass splitting. 

The remaining five conditions are taken from the scalar sector of the model. 
Unfortunately, the scalar channel in the region about $1$\ GeV became a 
long-standing problem of QCD. The abundance of meson resonances with $0^{++}$ 
quantum numbers shows that one can expect the presence of non-$q\bar q$ scalar
objects, like glueballs, hybrids, multiquark states and so forth 
\cite{Klempt:2007}. This creates known difficulties in the interpretation and 
classification of scalars. For instance, the numerical attempts to organize the 
$U(3)$ quark-antiquark nonet based on the light scalar mesons, $\sigma$ or 
$f_0(600),$ $a_0(980),$ $\kappa (850), f_0(980)$, in the framework of NJL-type 
models have failed (see, e.g. \cite{Weise:1990,Vogl:1990,Weise:1991,Volkov:1984,Volkov:1986,Osipov:2004b,Su:2007}). The reason is the ordering of the 
calculated spectrum which typically is $m_\sigma<m_{a_0}<m_\kappa <m_{f_0}$, as 
opposed to the empirical evidence: $m_\kappa<m_{a_0}\simeq m_{f_0}$. 

On the other hand, it is known that a unitarized nonrelativistic meson model 
can successfully describe the light scalar meson nonet as $\bar qq$ states 
with a meson-meson admixture \cite{Beveren:1986}. Another model which assumes 
the mixing of $q\bar q$-states with others, consisting of two quarks and two 
antiquarks, $q^2\bar q^2$ \cite{Jaffe:1977}, yields a possible description of 
the $0^{++}$ meson spectra as well \cite{Schechter:2008,Schechter:2009}. The 
well known model of Close and T\"ornqvist \cite{Close:2002} is also designed 
to describe two scalar nonets (above and below $1$\, GeV). The light scalar 
nonet below $1$\, GeV has a core made of $q^2\bar q^2$ states with a small
admixture of a $\bar qq$ component, rearranged asymptotically as 
meson-meson states. These successful solutions seemingly indicate on the 
importance of certain admixtures for the correct description of the light 
scalars. Our model contains such admixtures in the form of the appropriate 
effective multi-quark vertices with the asymptotic meson states described by 
the bosonized $\bar qq$ fields. We have found, that the quark mass 
dependent interactions can solve the problem of the light scalar spectrum and 
these masses can be understood in terms of spontaneous and explicit chiral 
symmetry breaking only. Indeed, one can easily fit the data: $m_\sigma =600$\ 
MeV, $m_{a_0}=980$\ MeV, $m_\kappa =850$\ MeV, $m_{f_0}=980$\ MeV with the input 
value $g_2=0$. In this case we obtain for the singlet-octet mixing angle 
$\theta_s$ roughly $\theta_s=19^\circ$.

\begin{table*}
\caption{Parameter sets of the model: $\hat m, m_s$, and $\Lambda$ are given in 
         MeV. The couplings have the following units: $[G]=$ GeV$^{-2}$, 
         $[\kappa ]=$ GeV$^{-5}$, $[g_1]=[g_2]=$ GeV$^{-8}$. We also show here 
         the values of constituent quark masses $\hat M$ and $M_s$ in MeV.}
\label{table-1}
\begin{tabular*}{\textwidth}{@{\extracolsep{\fill}}lrrrrrrrrl@{}}
\hline
Sets & \multicolumn{1}{c}{$\hat m$} 
     & \multicolumn{1}{c}{$m_s$}
     & \multicolumn{1}{c}{$\hat M$}                         
     & \multicolumn{1}{c}{$M_s$}                             
     & \multicolumn{1}{c}{$\Lambda$}    
     & \multicolumn{1}{c}{$G$}  
     & \multicolumn{1}{c}{$-\kappa$} 
     & \multicolumn{1}{c}{$g_1$}    
     & \multicolumn{1}{c}{$g_2$} \\ 
\hline
a  & 4.0* & 100* & 361 & 526 & 837  & 8.96  & 93.0   & 1534 &0* \\  
b  & 4.0* & 100* & 361 & 526 & 837  & 7.06  & 93.3   & 3420 &0* \\ 
\hline
\end{tabular*} 
\end{table*}  

\begin{table*}
\caption{Explicit symmetry breaking interaction couplings. The couplings have
the following units: $[\kappa_1]=$ GeV$^{-1}$, $[\kappa_2]=$ GeV$^{-3}$, 
$[g_3]=[g_4]=$ GeV$^{-6}$, $[g_5]=[g_6]=[g_7]=[g_8]=$ GeV$^{-4}$,
$[g_9]=[g_{10}]=$ GeV$^{-2}$.}
\label{table-2}
\begin{tabular*}{\textwidth}{@{\extracolsep{\fill}}lrrrrrrrrrl@{}}
\hline
Sets  & \multicolumn{1}{c}{$\kappa_1$}
      & \multicolumn{1}{c}{$\kappa_2$}    
      & \multicolumn{1}{c}{$-g_3$}  
      & \multicolumn{1}{c}{$-g_4$} 
      & \multicolumn{1}{c}{$g_5$} 
      & \multicolumn{1}{c}{$-g_6$}   
      & \multicolumn{1}{c}{$-g_7$} 
      & \multicolumn{1}{c}{$g_8$} 
      & \multicolumn{1}{c}{$g_9$}
      & \multicolumn{1}{c}{$g_{10}$} \\ 
\hline
a  &0* & 9.05  & 4967 & 661 & 192.2 & 1236  & 293  & 52.2  &0* &0*  \\  
b  &0* & 9.01  & 4990 & 653 & 192.5 & 1242  & 293  & 51.3  &0* &0*  \\ 
\hline
\end{tabular*}
\end{table*}   
    
To many readers our success with scalars may seem trivial: we have five 
parameters to fit five numbers. What is not trivial, however, is that the 
overall result of the fit is also in an agreement with phenomenological 
expectations. To compare, if we try instead to fit the second scalar nonet 
$f_0(1370), a_0(1450), K_0^*(1430), f_0(1500)$ with the same input, our attempt
fails. The best that we can do is the values $m_{f_0}=1220$\ MeV, $m_{a_0}=
1406$\ MeV, $m_{K^*_0}=1506$\ MeV, $m_{f_0'}=1786$\ MeV. However, even these 
unreasonable masses come out only together with the very large ratio 
$m_s/\hat m=36$ and phenomenologically unacceptable values for constituent 
quark masses $\hat M=631$\ MeV and $M_s=919$\ MeV. 

We obtain and understand the empirical mass assignment inside the light scalar 
nonet as a consequence of the quark-mass dependent interactions, i.e. as the 
result of some predominance of the explicit chiral symmetry breaking terms over
the dynamical chiral symmetry breaking ones for these states. Indeed, let us 
consider the difference 
\begin{eqnarray}
\label{a0-K}
   m_{a_0}^2-m_\kappa^2&\!=\!&2g^2\left(\frac{1}{H_{a_0}}-
   \frac{1}{H_\kappa}\right) \nonumber \\ 
                     &\!-\!&2(M_s+2\hat M)(M_s-\hat M). 
\end{eqnarray}
The sign of this expression is a result of the competition of two terms. In the 
chiral limit both of them are zero, since at $\hat\mu , \mu_s =0$ we obtain 
$\hat M=M_s$ and $H_{a_0}=H_\kappa$, for $H_{a_0}$ and $H_\kappa$ being positive. 
The splitting $H_\kappa >H_{a_0}$ is a necessary condition to get $m_{a_0}>
m_\kappa$. The following terms contribute to the difference
\begin{eqnarray}
\label{dif}
   H_\kappa -H_{a_0}&\!=\!& \kappa (h_s-\hat h)+2\kappa_2(\mu_s -\hat \mu )
           -g_2 (h_s^2+\hat hh_s -2\hat h^2) \nonumber \\
           &\!+\!&\frac{g_3}{2}\left(2\mu_s h_s + \mu_s\hat h + \hat\mu h_s
           -4\hat\mu\hat h\right)
           \nonumber \\
           &\!+\!&g_5\hat\mu (\mu_s -\hat\mu )+\frac{g_6}{2}
           \left(\mu_s^2-\hat\mu^2\right).
\end{eqnarray}
Accordingly, from this formula we deduce the ``anatomy'' of the successful
numerical fit:
\begin{eqnarray}
   m_{a_0}^2-m_\kappa^2&\!=\!&\left( [0.007]_\kappa + [0.076]_{\kappa_2}  
                        + [0]_{g_2} \right. \nonumber \\
                     &\!+\!&[0.832]_{g_3} + [0.003]_{g_5} +[-0.269]_{g_6} 
                     \nonumber  \\
                     &\!-\!&\left. [0.41]_M = 0.24\right)\ \mbox{GeV}^2,
\end{eqnarray}
where the contributions of terms with corresponding coupling (see eq. 
(\ref{dif})) are indicated in square brackets. The last number, marked by $M$, 
is the value of the last term from (\ref{a0-K}). It is a contribution due to  
the dynamical chiral symmetry breaking (in the presence of an explicit chiral 
symmetry breaking). One can see that the $g_3$-interaction is the main reason 
for the reverse ordering $m_{a_0}>m_\kappa$, the coupling $g_6$ being responsible
for the fine tuning of the result. 

Let us now show the result of our global fitting of the model parameters. We
collect them in two tables. Two sets (a) and (b) are shown. The difference is
the fitted value of the $\sigma$ mass: in (a) $m_{\sigma}=600$\ MeV, in (b) 
$m_{\sigma}=500$\ MeV. Table 1 contains the standard set of parameters, which 
are known from previous considerations. Their values are not much affected by 
the quark mass effects. Table 2 contains the couplings which are responsible 
for the explicit chiral symmetry breaking effects in the interactions. Note 
that these couplings almost do not change from set (a) to (b). We have already 
learned (as seen again in Table 1) that higher values of $g_1$ lead to the 
lower $\sigma$ mass \cite{Osipov:2006a}. This eight-quark interaction violates 
Zweig's rule, since it involves $q\bar q$ annihilation. The mixing angle 
$\theta_s$ is stable with respect to such changes, we obtain $\theta_s 
=19.4^\circ$ in case (a), and $\theta_s=18.9^\circ$ in case (b). The calculated 
values of quark condensates are the same for both sets: $-\langle\bar uu
\rangle^{\frac{1}{3}}=232$\ MeV, and $-\langle\bar ss\rangle^{\frac{1}{3}}=206$\ 
MeV. Our calculated values for constituent quark masses agree with the ones 
found in \cite{Georgi:1984,Weise:1990,Vogl:1990,Weise:1991}, showing their 
insensitivity to the new mass-dependent corrections.   

\section{Concluding remarks}

The purpose of this paper has been to take into account the quark masses at 
next to leading order in the expansion of the effective multi-quark Lagrangian
of the NJL-type. As a result a picture with some attractive new features 
has emerged. Let us summarize the details of such a picture. 

The main qualitative difference between our result and previous calculations is 
the possibility to fit the low lying pseudoscalar spectrum (the pseudo 
Goldstone $0^{-+}$ nonet) and weak decay constants of the pion and the kaon to 
perfect accuracy. The fitting of the $\eta\!-\!\eta'$ mass splitting together 
with the overall successful description of the whole set of low-energy 
characteristics is actually a solution for a long standing problem of NJL-type
models. We expect that with such modifications the model is getting more 
appropriate not only for studying low-energy meson physics, but also in 
studies of the ground state of hadronic matter in an environment, which is 
known to be very sensitive to quark mass effects. 

With a set of new quark-mass dependent interactions we are also capable to 
describe the spectrum of the light scalar nonet. From that one can conclude 
that both spectra can be understood on the basis of the dynamical and explicit 
chiral symmetry breaking only. The splitting inside the scalar nonet is 
determined by two competing contributions: first it is due to the explicit 
symmetry breaking (embodied in the stationary phase part of the bosonized 
Lagrangian), second it is due to the dynamical symmetry breaking (see the heat 
kernel part of the bosonized Lagrangian). It is the first type of contribution 
that changes the ordering inside the light scalar nonet, as compared to the 
standard approach.     

Our result for the scalar sector, being promising by itself, must be 
considered with some reservation. To report about a real success here, one 
should explain not only the mass spectrum of scalars, particularly the mass 
degeneracy of the $f_0(980)$ and $a_0(980)$ states (as we have done here), 
but answer some known challenges related with radiative decays of these states. 
Work in this direction is in progress.

{\bf Acknowledgements}    
This work has been supported by the Funda\c{c}\~ao para a Ci\^encia e 
Tecnologia, project: CERN/FP/116334/2010, developed under the iniciative QREN,
financed by UE/FEDER through COMPETE - Programa Operacional Factores de 
Competitividade. This research is part of the EU Research Infrastructure 
Integrating Activity Study of Strongly Interacting Matter (HadronPhysics3)
under the 7th Framework Programme of EU, Grant Agreement No. 283286. 


\end{document}